# Isochronous Data Transmission With Rates Close to Channel Capacity.

Alexander Zhdanov
Voronezh, Russia
Email a-zhdan@vmail.ru

*Abstract* **The existing ARQ schemes (including a hybrid ARQ) have a throughput depending on packet error probability. In this paper we describe a strategy for delay tolerant applications which provide a constant throughput until the algorithm robustness criterion is not failed. The algorithm robustness criterion is applied to find the optimum size of the retransmission block in the assumption of the small changes of coding rate within the rate compatible codes family.**

## I. Introduction

As it was known from earlier work of Elias [1] and Wozencraft [2] reliable communication over noisy channels could be provided by using two main methods: forward error correction coding (FEC) and automatic repeat request (ARQ). The classical ARQ includes full retransmission of erroneous packets by the receiver request transmitted over the feedback channel. Throughout the paper we assume error free feedback channel where each decoding attempt is acknowledged (negative acknowledged) by an ACK (NACK) symbol. The performance of such a communication system is defined in terms of throughput, delay and undetected error probability. Throughput is defined as a ratio of accepted and really transmitted bits; delay is defined as a time interval between the time of actual packet acceptations and the arrival time in a system without noise. To improve the throughput if delay is not significant one can use a hybrid ARQ (HARQ) method [3] which includes retransmission and properly combining of different coded symbols relating to the same message. At the receiver side the previous and new coded bits are properly combined and decoded together. The HARQ needs to store the coded symbols in soft decisions in a buffer. The method of incremental redundancy coding [4] or HARQ type II includes the following: using a systematic FEC code, separating redundant parity check symbols into subblocks that may be not necessarily equal length and transmitting the subblock if and only if previous decoding attempt has failed. This method was developed and practically adopted for convolutional codes in [5, 6]. In [7] the Hagenauer idea of template puncturing of mother convolutional code was applied for parallel convolutional turbo codes (PCTC) to obtain a rate compatible turbo codes (RCPTC) family. Since a number of papers were published where some puncturing patterns have been proposed but theoretical analysis of puncturing code ensembles is still a hard task. It requires analytical prediction of word error probabilities (WER) of punctured codes with further optimization of the retransmission block length. For example, in [8] the union bounding technique is used to estimate WER but such a method is valid only in high SNR region. Note, for the data transmission with allowable delay the optimum value of the target WER of the code $C$ could be equal to $P_W^C \approx 0.3 \div 0.5$. The closest results could be obtained by tangential sphere packing bound (TSB) [9] but such a technique requires considerable computational efforts and code spectrum estimation. The obtained result slightly differs from simulation results of iterative decoding of the turbo code in a low SNR region but this difference of approximately $10\% \div 20\%$ is important for the throughput analysis. We refer the reader to [10] for detailed review of the known bounding technique. The famous random coding bound [11] where the minimum WER of linear block code for the rates below the channel capacity for any block length $N$ and any number of codewords $M = \exp(NR_{nat})$, where $R_{nat}$ is a code rate in nats is bounded by

$$P_W \le \exp\left(-NE(R_{nat})\right) \quad (1)$$

where $E(R_{nat})$ is a random coding exponent , a positive function in $0 \le R_{nat} \le C$ defined as follows:

$$E = \begin{cases} R_0 - R_{nat},\ 0 \le R_{nat} \le R_{crit} \\ E_{sp}\left(R_{nat}\right), R_{crit} \le R_{nat} \le C \end{cases} \quad (2)$$

where $R_0$ is a cutoff rate, $R_{crit}$ is a critical rate and $E_{sp}$ is defined in the parametric function $\rho$ as follows

$$\begin{cases} E_{sp} = \ln(\beta) + (1+\rho)(1-\beta) \\ R_{nat} = \ln\left(\beta + A/(1+\rho)\right) \end{cases},\ 0 \le \rho \le 1 \quad (3)$$

where $A$ - power signal noise ratio per coded symbol and $\beta$ defined as follows:

$$\beta = \frac{1}{2} \times \left\{ 1 - \frac{A}{1+\rho} + \sqrt{\left(1 - \frac{A}{1+\rho}\right)^2 + \frac{4A}{(1+\rho)^2}} \right\}. \quad (4)$$

Another way of definition of the error exponent $E$ is given by

$$E(R) = \max_{0 \le \rho \le 1}\left(-\rho R_{nat} + E_0(\rho)\right). \quad (5)$$

This bound is useless for direct application due to imperfectness of the known codes but in this paper we will use the random coding bound to predict conditional retransmission probabilities. Let us consider the isochronous data transmission system with decision feedback where each packet has the same size and is transmitted within the defined time interval. If we assume an error free channel each packet should be delivered without any delay in the own time transmission interval. However what minimal coding redundancy do we really need to keep the same throughput in presence of noise if we introduce a supplemental ARQ channel with the fixed number of the transmitted symbols? In such case the delay tolerance of the data traffic is required. We assume that the data traffic is delay tolerant if the number T exists for any packet of this traffic and for any probability $\varepsilon_0 > 0$, such that the packet will be delivered with probability $1-\varepsilon_0$ within $T$ normal transmission intervals. It is notable that if such algorithm works the delay of a single packet does not impact on the delay of other packets in opposite to the classical ARQ methods and the throughput does not depend on the packet error probability if that error is detectable. Such a scheme is useful, for example, for multicast transmission with HARQ protocols.

The algorithms where puncturing performed to provide free places for incremental redundancy retransmission are known from [15,16]. The algorithm [16] will be discussed in this paper as Algorithm A. The delay performance of such algorithm is obtained in [17]. The aim of this paper is to provide a theoretical framework for throughput analysis of the proposed scheme and to find the optimum size of the retransmission block in the assumption of the small changes of coding rate within the RCPTC family.

## II. Incremental Redundancy and Isochronous Transmission

The data transmission system is scheduled that the ARQ channel always exists and the number of transmitted symbols per packet transmission interval is fixed and equal to the sum of symbols transmitted per normal and ARQ channels. Let $C_{-1} \subset C_0 \subset C_1 \subset C_2 \subset \ldots \subset C_m$ be a rate compatible code family. The index means relative number of optional coding groups. The model of the proposed HARQ-II system is given by Fig. 1. The CRC encoder adds $\Delta K$ parity bits to $K$ information bits. The $K_0 = K + \Delta K$ bits are encoded to obtain $N_{-1}$ coded bits for normal transmission and $m$ groups of size $\Delta N$ bits for HARQ-II retransmissions. Let $P_W^{C_i}$ be WER of the code $C_i$. The obtained coded packet of the code $C_{-1}$ is transmitted over the normal channel. At the receiver end the packet is decoded and CRC check is passed. The decoding results are immediately acknowledged. For simplicity we do not take into account a feedback delay. If sending NACK the HARQ-II retransmission is sent over the ARQ channel to extend the codeword of the code $C_{-1}$ stored in the FCFS queue to the codeword of the code $C_0$. In parallel the next codeword of the code $C_{-1}$ is transmitted over the normal channel. At the receiver end both packets are decoded but the stored packet is decoded with arriving HARQ-II retransmission.

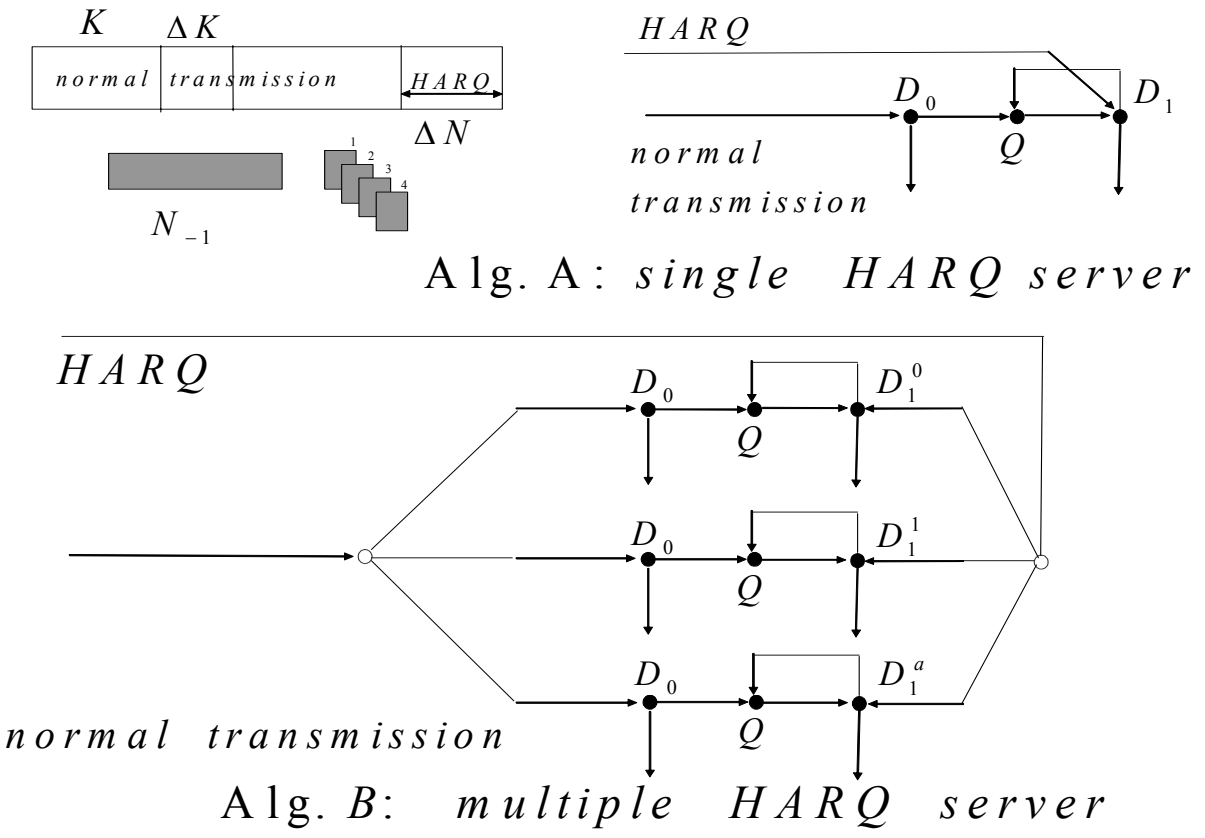

Fig. 1 HARQ-II for isochronous data transmission

If the new packet does not pass the CRC check the packet is directed to the queue. If the stored packet is decoded successfully this packet is removed from the queue. The packet could not be removed from the queue until it successfully passes the CRC check. If there are no packets in the queue (both packets are decoded successfully) the consequent packet could be transmitted with the additional coding group transmitted over the ARQ channel simultaneously with normal transmission forming the codeword of the code $C_0$ immediately. That is a method with a single HARQ server (Algorithm A). The Algorithm B is a multiple HARQ server system where the HARQ channel is shared between HARQ retransmissions for different $a$ packets. The arrived packets are directed to one of the single HARQ servers by the deterministic law known at the transmitter side. In such a way the transmitter could form multiple subblocks for all HARQ servers by the deterministic law known at the receiver side.

The proposed schemes are shown in Fig.1 where the module $D$ contains an FEC decoder and a CRC decoder. The packet arriving at the module $D$ is decoded correctly with probability $1-P_W^{C_i}$ or decoded with errors with probability $P_W^{C_i}$ detected by CRC. In the latter case the packet goes to the queue $Q$. This algorithm is robust if the queue is a stationary system. The criterion and the model are based on the known error probabilities $P_W^{C_i}$. The criterion is quite simple if we define basic properties of the error probability set $P_W^{C_i}$. First, $P_W^{C_i} > P_W^{C_j}$ when $i < j$ and second, $\lim_{m\to\infty} P_W^{C_m} = 0$. In such a way the arrival rate is defined as $\lambda = P_W^{C_{-1}}$ (if the queue is not empty). The average departure time from the queue after $m$ retransmission attempts is equal to

$$t_{out} = \frac{\sum_{i=0}^{m}\left(P_W^{C_{i-1}} - P_W^{C_i}\right)(i+1)}{\left(P_W^{C_{-1}} - P_W^{C_m}\right)}. \quad (6)$$

Then we rush $m \to \infty$ and the service rate should be written as

$$\mu = \frac{1}{t_{out}} = \frac{P_W^{C_{-1}}}{\sum_{i=0}^{\infty}\left(P_W^{C_{i-1}} - P_W^{C_i}\right)(i+1)}. \quad (7)$$

The queue is stable if $\mu > \lambda$ or

$$\sum_{i=0}^{\infty}\left(P_W^{C_{i-1}} - P_W^{C_i}\right)(i+1) = \sum_{i=-1}^{\infty} P_W^{C_i} < 1. \quad (8)$$

For the multiple HARQ server, (8) could be rewritten as

$$\sum_{i=-a}^{\infty} P_W^{C_i} < a. \quad (9)$$

There exist several classes of error detection codes where the undetected error probability is not greater than $2^{-\Delta K}$ as it was shown in [12]. Since error detection is independent from error correction, the undetected error probability should be upper bounded as $m' \times 2^{-\Delta K}$ where $m' = m+1$ is a number of decoding attempts including the decoding of the normal transmission. The average number of decoding attempts is less than or equal to the number of the parallel operating decoders $a+1$. If two decoders are used in parallel the undetected error probability increases only two times so one bit is enough to compensate the losses. For any practically purpose we can choose redundancy of the error detection code $\Delta K$ as

$$\Delta K \geq -\log_2(\varepsilon_1) + \log_2(a+1) \quad (10)$$

where $\varepsilon_1$ is a target undetected error probability. The relative throughput of this system is equal to 1 packet per time transmission interval. Throughput of the system with full retransmission is equal to $1 - P_W^{C_0}$. Thus we can formulate the criterion of optimality: minimize the SNR keeping constraints (8) or (9) to obtain the maximum value of $P_W^{C_0}$.

## III. The Series of Word Error Probabilities

However to solve this optimization task we need at least to know dependencies in the series $P_W^{C_i}$ .Let $E(R) \triangleq E^{C_i}(R)$ be an error exponent of the rate compatible code (RCC) family $C_i$ such that

$$P_W^{C_i} = \exp\left(-(N + i\Delta N) \times E(R_{nat})\right) \quad (11)$$

Let us consider redundancy $z_i = 1 / R_{nat}^{C_i}$ in a way that WER is defined as

$$P_W^{C_i} = \exp\left(-K_0 \times \ln(2) \times z_i \times E(z_i)\right). \quad (12)$$

Redundancy $z_i$ of the rate compatible codes $C_i$ rises uniformly, thus $z_{i+1} = z_i + \Delta z$ . Let $K_1 = K_0 \ln(2)$ and choosing $z$ such that $z_{i+1} > z > z_i$ we have

$$\lim_{\Delta z \to 0} \frac{P_W^{C_i}}{P_W^{C_{i+1}}} = \frac{\exp\left(-K_0 z_i E(z_i)\right)}{\exp\left(-K_0 z_{i+1} E(z_{i+1})\right)} = \exp\left(-K_0 \left(zE(z)\right)'\right). \quad (13)$$

We use the property of function $E_0$ which is given by

$$E = E_0(\rho) - \rho\frac{\partial E_0(\rho)}{\partial \rho}, \quad \frac{\partial E_0(\rho)}{\partial \rho} = R_{nat}. \quad (14)$$

Replacing the function $E$ by function $E_0$ we obtain the following:

$$\left(zE(z)\right)' = \left(E_0(z) + \frac{\partial \rho}{\partial z}\left(\frac{\partial E_0(z)}{\partial \rho} z - 1\right)\right)\Delta z. \quad (15)$$

It is easy to see that

$$\frac{\partial E_0(z)}{\partial \rho} z - 1 = Rz - 1 = 0 \quad (16)$$

then

$$\lim_{\Delta z \to 0} \frac{P_W^{C_i}}{P_W^{C_{i+1}}} = \exp\left(-K_1\left(E_0(z)\Delta z\right)\right). \quad (17)$$

Let $E_0^1 = R_0$ when $0 \leq R_{nat} \leq R_{crit}$ and $E_0^1 = E_0$ when $R_{crit} \leq R_{nat} \leq C$. The functions $E$, $E_0^1$ are shown in Fig. 2 with the parameter $A = 0.3453$.

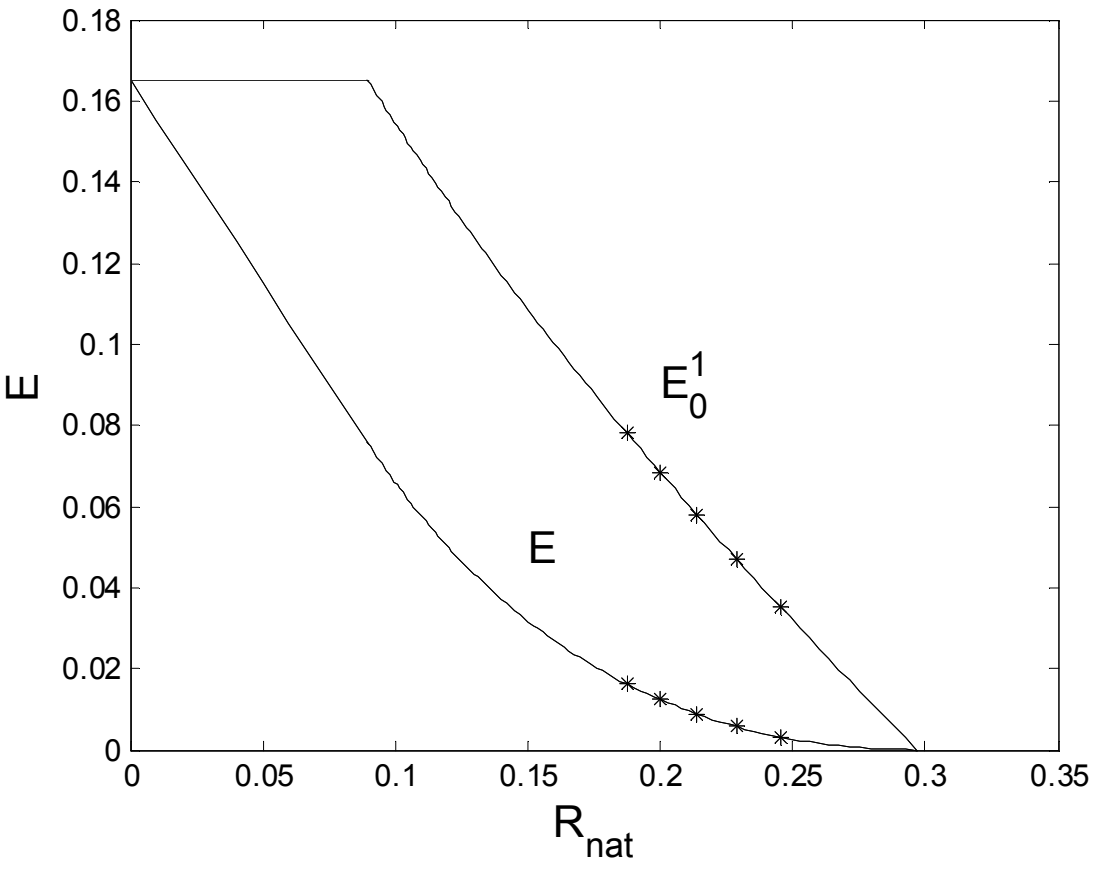

Fig. 2 The error exponent and RCC family

The points corresponding to the RCC family are also shown in Fig. 2. The binary rates are given as a relation of the number of information bits to the number of coded bits equal to $\frac{16}{47}, \frac{16}{48}, \frac{16}{49}, \frac{16}{50}, \frac{16}{51}$. Considering light changes of the rate we propose to approximate the function $E_0$ by the straight line and choose $P_{-1} = P_W^{C_{-1}}, P_0 = P_W^{C_0}, P_1 > P_W^{C_1}, \dots, P_n > P_W^{C_n}$, where the estimates $P_{-1}, P_0, P_1, \dots, P_n$ are defined as follows:

$$\begin{aligned} &P_{-1} \\ &P_0 = P_{-1} \times (h) \\ &P_1 = P_{-1} \times (h) \times (hg) \\ &P_2 = P_{-1} \times (h) \times (hg) \times (hg^2) \\ &P_n = P_{-1} \times (h^n) \times \left( g^{\frac{n(n-1)}{2}} \right) \end{aligned} \quad (18)$$

where $g = 1$ if $0 \le R_{nat} \le R_{crit}$ and $g < 1$ if $R_{crit} \le R_{nat} \le C$. We assume that

$$\lim_{\Delta z \to 0} \left( \frac{P_W^{C_{i+2}}}{P_W^{C_{i+1}}} : \frac{P_W^{C_{i+1}}}{P_W^{C_i}} \right) = \exp\left(-K_1\left(\Delta E_0^1(z)\Delta z\right)\right) \approx g. \quad (19)$$

## IV. THROUGHPUT ANALYSIS

Let us start with a geometric series case when $g = 1$. The criterion of stationary (8) could be written as:

$$\sum_{i=0}^{\infty} P_{-1}(h^i - h^{i+1})(i+1) \triangleq k_0 < 1. \quad (20)$$

Let $k_1(n) = \sum_{i=0}^{n} h^i (i+1)$, $k_2(n) = \sum_{i=0}^{n} h^{i+1}(i+1)$. It is easy to see that $k_1(n) = hk_2(n)$ and

$$\lim_{n \to \infty} (1-h) k_1(n) P_{-1} = k_0 < 1. \quad (21)$$

The sum of the first series is equal to

$$k_1(n) = \lim_{n\to\infty} \sum_{i=0}^{n} h^i (i+1) = \left[1 + 2h + \dots + nh^{n-1}\right] = \frac{1}{(1-h)^2}. \quad (22)$$

Thus, substituting (22) to (21) we have

$$\frac{P_{-1}}{(1-h)} = k_0 < 1. \quad (23)$$

To estimate the possible gain comparing with the classic ARQ we need to maximize $P_0 = hP_{-1} = k_0(1-h)h$ as a function of $h$.

It is obvious that the maximum is achieved when $h = 1/2$, $P_{-1} = k_0/2$, $P_0 = k_0/4$.

Considering the case of the multiple HARQ server one can note that the single server solution is valid if we make the following substitution $P_{-1}^* = P_{-1}/a$; $h^* = h^{1/a}$. And we need to maximize

$$P_a^* = k_0 \left(1 - h^{\frac{1}{a}}\right)\left(h^{\frac{1}{t}}\right)^t = k_0 \left(1 - h^{\frac{1}{a}}\right) h. \quad (24)$$

The maximum is achieved when $\left(\frac{a}{a+1}\right)^a = h$. So, the maximum possible throughput gain of the multiple HARQ server comparing to the classical ARQ when $a \to \infty$ is equal to $\lim_{a\to\infty}\left(\frac{a}{a+1}\right)^a = \frac{1}{e} = 0.3679$. But the required CRC redundancy $\Delta K$ also tends to infinity according to (10). So, the algorithm A can win 25% of throughput and the algorithm B can win up to 36,79% of throughput compared to the classical ARQ data transmission system if the geometric series is formed by WER. But its true only if the error exponent of such code family is a straight line. If we want to optimize the performance such HARQ-II system based on RCPTC we shall note, that in spite the turbo code is asymptotically bad and its performance could not be represented by error exponent the approximation (18) is valid until the error floor is not reached. According to [13] we assume that the word error probability is $O(1)$ for PCTC with the two branches in the error floor region. We propose numerical optimization of the retransmission block length $\Delta N$ using obtaining results as a basis solution. We assume that we can form a retransmission block of arbitrary length and we can to approximate the throughput loss due to error floor as a function inversely proportional to $\Delta N$.

1. Choose $\Delta N_{bas}$, and SNR such that $\hat{P}_{-1} \simeq 0.5$ and $\hat{P}_0 \simeq 0.25$ where $\hat{P}_i$ is WER obtained by simulation
2. Estimate the values of $\hat{h} = \hat{P}_0 / \hat{P}_{-1}$ and $\hat{g} = \left(\hat{P}_{-1} \times \hat{P}_1\right) / \left(\hat{P}_0^{\,2}\right)$ and estimate the value of $\theta = \sum_{i=j}^{i=\infty} \hat{P}_i$ where $j$ is a minimum number such that $\hat{P}_j < 0.01$.
3. Find $r' = \mathrm{arqmin}\left[\phi(r) + \theta/r\right]$ where $\phi(r) = \hat{P}_0 \sum_{n=0}^{n=\infty} \left(\hat{h}^{rn}\right) \times \left(\hat{g}^{\left(r^2\right)\frac{n(n+1)}{2}}\right) + \frac{\hat{P}_0 \hat{g}^{(r-1)}}{\hat{h}^r}$, $r > 0$
4. Set the value of retransmission block length $\Delta N = r' \Delta N_{bas}$ for given block length $N_0$;

## V. EXAMPLE AND SIMULATION RESULTS

We have simulated the algorithm A with PCTC. We use the 3GPP turbo code [14] as the code $C_0$ with $K = 2048$ and the binary rate $R_0 = 1/3$ which could be extendable to the rate 1/5 by the method defined in [7] with further repetition of a systematic part for the Chase combining [5]. We have used the same random transmission assignment as defined in [8] to shape retransmission blocks of arbitrary length. We have generated the rate compatible code family as it was shown in Fig. 2. Throughput simulation results are given in Fig. 3. The results of $\Delta N$ optimization with simulated points obtained as $\sum_{i=-1}^{i=\infty} \widehat{P}_i$ are given in Fig. 4.

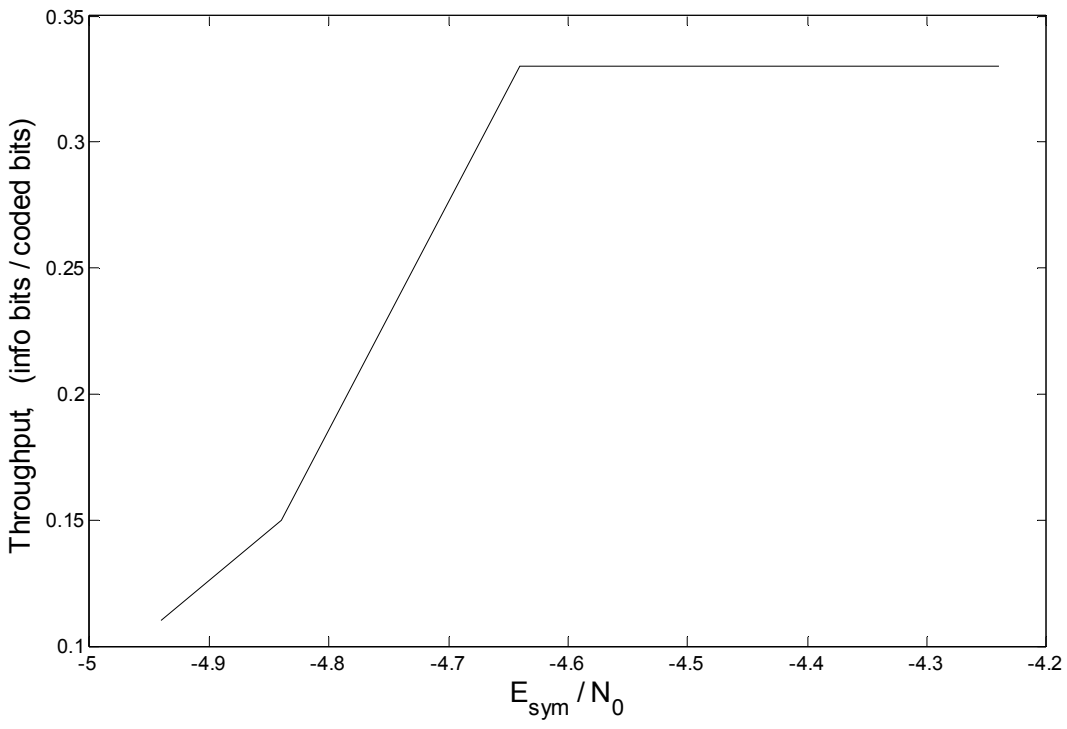

Fig. 3 Throughput versus $E_{sym}/N_0$ for $\Delta N = 142$

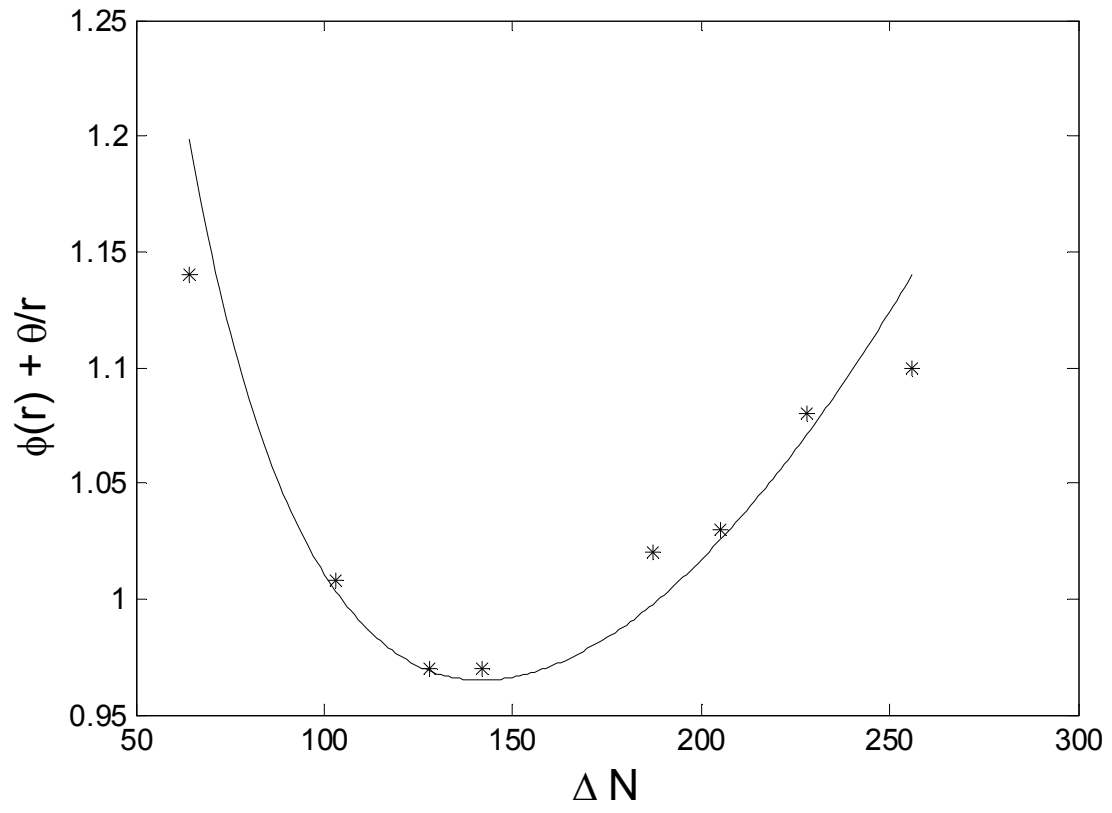

Fig. 4 $\Delta N$ optimization for $E_{sym}/N_0 = -4.64\ db$

## VI. Conclusion

The key feature of the proposed strategy is in using multiple decoding within the time transmission interval so the mathematical expectation of the number of the decoded packets is not less than 1. When the criterion (8) or (9) is satisfied, throughput $T$ of such an HARQ system does not depend on retransmission WER's. The maximum achieved WER when system is still stable in the critical point $E_{sym}/N_0 = -4.64\ db$ is equal to $\widehat{P}_0 = 0.2582$. Payment for the improved throughput is non-uniform packet delivery and extended feedback channel with limited rising of the undetected error probability. The feedback channel has to be extended at least from 1 bit per time transmission interval to 2 bits per time transmission interval. The required extension of the error correction code is defined by (10). The delay of such scheme could be obtained by the method from [17] where the Markov chain model is developed. The theoretically found optimum size of $\Delta N = 142$ is coincide with simulation results.